\documentclass[useAMS,usenatbib,usegraphicx]{mn2e}

\providecommand{\aap}{A\&A}
\providecommand{\aaps}{A\&AS}
\providecommand{\aj}{AJ}
\providecommand{\apj}{ApJ}
\providecommand{\apjl}{ApJ}

\providecommand{\mnras}{MNRAS}

\providecommand{\nat}{Nat}

\title[Spiral structure and the CMa debate]{Spiral structure of the
  Third Galactic Quadrant and the solution to the Canis Major debate}
\author[A. Moitinho et al.]{
A. Moitinho,$^{1}$\thanks{E-mail:andre@oal.ul.pt (AM)}
R. A. V\'azquez,$^{2}$
G. Carraro,$^{3}$\thanks{on leave from Dipartimento di
  Astronomia, Universita' di Padova, Italy.}
G. Baume,$^{2}$ 
E. E. Giorgi$^{2}$ and \newauthor
W. Lyra$^{4}$
\\
$^{1}$CAAUL, Observat\'orio Astron\'omico de Lisboa, Tapada da Ajuda,
  1349-018 Lisboa, Portugal
\\
$^{2}$Facultad de Ciencias Astron\'omicas y Geof\'{\i}sicas de la
  UNLP, IALP-CONICET, Paseo del Bosque s/n 1900, La Plata, Argentina
\\
$^{3}$ANDES Fellow,  Departamento de Astronomia, Universidad de Chile,
Chile, and Astronomy Department, Yale University, USA
\\
$^{4}$Department of Astronomy and Space Physics, Uppsala Astronomical
  Observatory, Box 515, 751\,20 Uppsala, Sweden}

\begin{document}

\date{Accepted ?. Received ?; in original form ?}

\pagerange{\pageref{firstpage}--\pageref{lastpage}} \pubyear{2006}

\maketitle

\label{firstpage}

\begin{abstract}
  With the discovery of the Sagittarius dwarf spheroidal
  \citep{Ibata1994}, a galaxy caught in the process of merging with
  the Milky Way, the hunt for other such accretion events has become a
  very active field of astrophysical research. The identification of a
  stellar ring-like structure in Monoceros, spanning more than 100
  degrees \citep{Newberg2002}, and the detection of an overdensity of
  stars in the direction of the constellation of Canis Major
  \citep[CMa;][]{Martin2004a}, apparently associated to the ring, has
  led to the widespread belief that a second galaxy being cannibalised
  by the Milky Way had been found.  In this scenario, the overdensity
  would be the remaining core of the disrupted galaxy and the ring
  would be the tidal debris left behind.  However, unlike the
  Sagittarius dwarf, which is well below the Galactic plane and whose
  orbit, and thus tidal tail, is nearly perpendicular to the plane of
  the Milky Way, the putative CMa galaxy and ring are nearly co-planar
  with the Galactic disk. This severely complicates the interpretation
  of observations.  In this letter, we show that our new description
  of the Milky Way leads to a completely different picture. We argue
  that the Norma-Cygnus spiral arm defines a distant stellar ring
  crossing Monoceros and the overdensity is simply a projection effect
  of looking along the nearby local arm.  Our perspective sheds new
  light on a very poorly known region, the third Galactic quadrant
  (3GQ), where CMa is located.
\end{abstract}


\section{Introduction}
The announced detection of a galaxy in CMa \citep{Martin2004a} centred
at Galactic coordinates l = 240$^{o}$, b = -8$^{o}$ and at a distance
of around 8 kpc from the Sun \citep{Martin2004a,MartinezDelgado2005},
has produced considerable excitement reaching well beyond the
astrophysical community.  Independently of how fascinating the idea,
the CMa galaxy scenario can be used to address several important
astrophysical questions:  it is the closest galaxy detected so far,
it can be used for a detailed study of the merging process. 
In having an orbit that
is nearly co-planar with the Galactic disc, it can contribute to
build up the thick disc thus favouring models of galaxy accretion as
the origin of this still poorly understood component. Finally, it
would bring the number of observed nearby low mass satellites of the
Milky Way closer to that expected from cosmological simulations
\citep{Klypin1999} of galaxy assembly.

If the presence of a galaxy so close to the Sun offers these
unique opportunities, it also requires a detailed knowledge of the
structure of the Milky Way to disentangle and understand the complex
interplay between both systems. Unfortunately, little attention has
been paid to the 3GQ in the past and apart from the presence of the
Galactic warp, little is known about its structure. In particular,
spiral structure has not been clearly mapped \citep{Russeil2003}.

\begin{figure*}
\includegraphics[width=\textwidth]{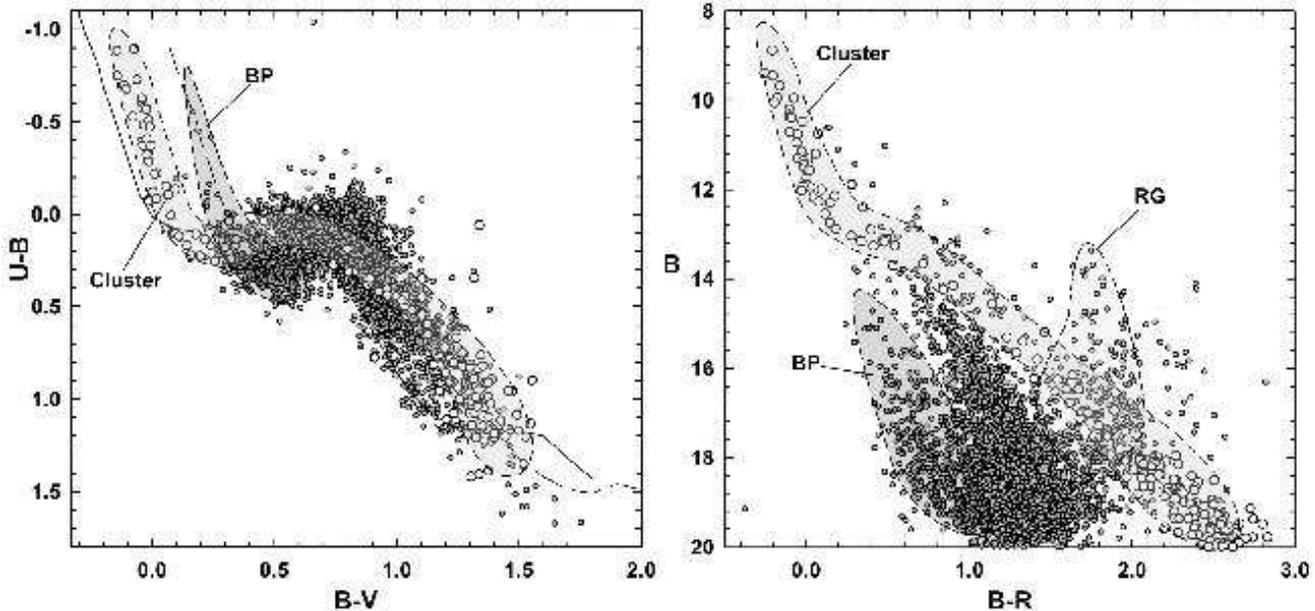}
\caption{Two-colour (TCD; left) and colour-magnitude (CMD; right)
  diagrams of a $9\times 9$ field around the open star cluster NGC
  2362 (l=238.18$^o$,b=-5.55$^o$). Shaded areas roughly separate the
  regions occupied by three stellar components: members of the cluster
  NGC 2362 -also shown with large open circles; the blue plume stars,
  BP, and the red giant stars, RG. The RG region is not shaded on the
  TCD to avoid confusion. Small grey filled circles indicate the
  galactic field dwarf population. For guidance, we have superposed on
  the TCD the intrinsic locus -continuous curve- for dwarf stars
  \citep{Schmidt-kaler1982} and the same curve shifted to account for
  the effect of reddening -dashed curve- to fit the average BP stars.
  It can be seen on the TCD that the BP includes stars with spectral
  types B5-A5, meaning unambiguously that it is a young population
  less than 100 Myr old. Unlike the CMD where BP stars, cluster
  members and the field population are well detached, the TCD is
  entangled for cluster members and BP stars of late B- and early
  A-types. In the CMD, no obvious overdensity is seen in the RG zone
  (expected position for the red clump of a 4-10 Gyr galaxy at a
  distance of 8 kpc).  }
\label{fig:phot}
\end{figure*}

Apart from the ring and the overdensity, deep colour-magnitude
diagrams (CMDs) have been considered to provide additional evidence
supporting the reality of the CMa galaxy. By comparing CMDs with
stellar evolution models, studies have found the CMa galaxy to be at
distance of 8 kpc and to have an age of 4-10 Gyr
\citep{Bellazzini2004,MartinezDelgado2005}. Although the CMDs do not
exhibit clear post main-sequence signatures expected for a 4-10 Gyr
population \citep{MartinezDelgado2005} (red clump or red giant branch,
horizontal branch, RR-Lyrae), a distinctive feature, popularised as
the Blue Plume (BP; see Fig.~\ref{fig:phot}) has been taken as strong
evidence for the existence of the CMa galaxy
\citep{Bellazzini2004,Dinescu2005,MartinezDelgado2005}.  The BP has
been modelled and interpreted as the last burst of star formation in
that galaxy 1-2 Gyr ago (which, given that it is a 1-2 Gyr population,
should also have a red clump). This piece of evidence has been taken
to be an unambiguous indicator of the reality of CMa given that it
does not correspond to any known Galactic component.  Additionally,
the narrowness of the BP, which is indicative of a small distance
spread, has been taken as evidence for a compact, possibly bound
system \citep{MartinezDelgado2005}.  The absence of a clear post
main-sequence is ascribed to heavy contamination by the Galactic field
population.

We have recently found \citep{Carraro2005c} that the BP is actually
composed of a young population less than 100 Myr old.  Except for the
cluster sequence which is well detached from the rest of the stars,
the CMD of the field around the open cluster NGC~2362 shown in
Fig.~\ref{fig:phot} is identical to the one obtained in another study
\citep{MartinezDelgado2005} at l=240$^{o}$, b=-8$^{o}$, not far from
NGC~2362, and used as proof for the existence of the CMa galaxy.  In
particular, the BPs have the same position, shape and extension, but
we find the BP in Fig.~\ref{fig:phot} to be much younger and farther
away, at 10.8 kpc.

The enormous discrepancy between our study and the others deserves a
closer look. The photometric analysis presented in other studies were
exclusively limited to CMDs. It is well known that the determination
of fundamental parameters - reddening, distance, age and metallicity -
is affected by a number of degeneracies (that are readily admitted by
the authors) when using a single CMD: different sets of solutions are
equally acceptable within observational errors.  Those studies have
relied on complex modelling of the observed CMDs including the
expected Galactic field in that direction. Trial and error changes to
the galaxy's fundamental parameters, number of stars and star
formation history are made until the synthetic and observed diagrams
are considered to match. But even given the rigour of the modelling,
the degeneracies persist.  Furthermore, the descriptions of the
expected Galactic field are also synthetic.  Because in these models
the contributions from the halo, thick and thin discs do not reproduce
the BP, it has been argued that the BP does not correspond to any
known Galactic component. These models do not include spiral arms in
the region of the 3GQ under analysis.

Our results are based on UBVRI five band photometry
which {\it does} allow the determination of unique solutions. The reason is
that, apart from CMDs, two-colour diagrams (TCDs), which are distance
independent, are also used in the analysis. Moreover, when TCDs are
built using U band data (not used in the other studies), reddening and
spectral types of early type stars can be uniquely derived. Furthermore,
metallicity does not affect the colours of these stars significantly.
Hence the only unknowns that remain in the CMD analysis are distance and
age, which can also be uniquely derived provided that the photometry is
deep enough so that a population sequence appears with a well defined
morphology. 
Briefly, multi-band photometry including U measurements provides direct
determinations of reddening and spectral type whereas single CMD
analysis does not.

Young open clusters have long been recognised as privileged spiral
tracers \citep{Becker1970}. Their distances can be better
determined than those of individual stars and their youth keeps them
close to the spiral arms where they were born.  Over the last 10 years
we have collected observations resulting in a unique data set of
stellar photometry in the fields of many open star clusters
\citep{Moitinho2001a,Moitinho2001b,Carraro2003,Baume2004,Moitinho2006a}.
A sample of 61 open clusters has been obtained with the goal of
tracing the detailed structure of the Galactic disk in the 3GQ, which
we present and discuss in this letter.

\section{Results}
The clusters' basic parameters (reddening, distance and age) have been
derived via Zero Age Main Sequence (ZAMS) and isochrone fits to the
cluster sequences in different photometric diagrams. This is a
standard and solid method that has been used in open cluster studies
for several decades.  

\begin{figure}
\includegraphics[width=\columnwidth]{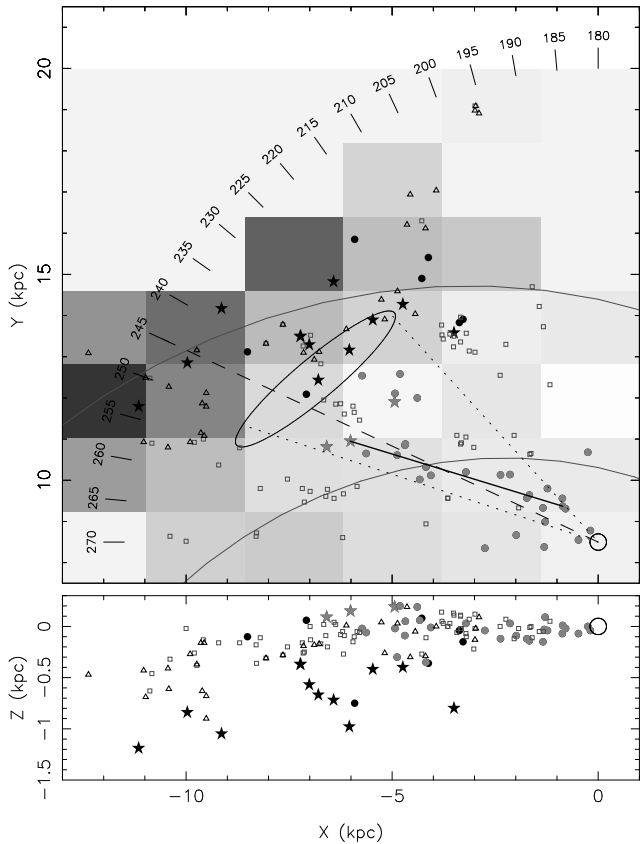}
\caption{
Distribution of young star clusters, BPs and massive CO clouds in the
third quadrant of the Milky Way. Clusters are depicted as filled
circles, BPs as stars. Darker symbols indicate the populations
associated to the Norma-Cygnus (Outer) arm. CO clouds are plotted as
empty triangles (newer data) and squares (older data). The coordinate
system is right handed with its origin at the Galactic centre. Y
indicates the direction of the Sun, Z points toward the north Galactic
pole and X grows in the direction of Galactic rotation at the position
of the Sun. The Sun is marked as a larger circle at X=0, Y=8.5, Z=0
kpc. The upper panel provides a view of the Galactic disc as seen from
above. A longitude scale is also provided. The grid of grey squares
illustrates the height of the disc with respect to the formal Galactic
plane ($b=0^o$): lighter tones are closer to the plane and darker
tones are deeper below. The tone scale is linear and each tone
corresponds to the average Z in the cell. A clear picture of the
Galactic warp is easily seen. The two curves that cross the panel are
model \citep{Vallee2005} extrapolations of the Outer and Perseus arms
and the solid straight line sketches the local arm. The position and
extent of the Canis Major overdensity are indicated by a large
ellipse. The dashed line is the line of sight mostly dominated by the
local arm and the dotted lines mark the range where the contribution
of the local arm appears to be significant. Lower panel: X-Z
projection. The signature of the warp is again prominent. It is also
readily visible how the the molecular clouds closely follow the
stellar distribution.
}
\label{fig:XYZ}
\end{figure}

From the 61 clusters in our sample, 25 were determined to be younger
than 100 million years. These are listed in Table~\ref{tab:pars} and
plotted in Fig.~\ref{fig:XYZ} which represents the third quadrant of
the Galactic plane seen from above.  Also plotted, are the BPs
detected in the backgrounds of several clusters.  A strip about 1.5
kpc wide, extending from l=210$^o$ to l=260$^o$, spanning distances
between 6 and 11 kpc depending on the line of sight, can be seen at
the position expected for the Outer (Norma-Cygnus) arm.  This strip is
mainly composed of BP detections, but also contains a few clusters.
On the X-Z projection, the putative outer arm members remain close to
the formal Galactic plane (b=0$^o$) up to l=220$^o$ where the spiral
arm starts descending, reaching around 1 kpc below the plane at
l=240-250$^o$. The bending of the arm is a clear signature of the
Galactic stellar warp.  This new optical detection of the Norma-Cygnus
arm, extending from l=210$^o$ to l=260$^o$, confirms our previous
interpretation of the BP as a spiral tracer and our optical detection
of the outer arm based on fewer points \citep{Carraro2005c}.  That the
BP appears so tight in the CMDs is then a natural consequence of the
arm's limited depth along each line of sight and is not, in this case,
a signature of a compact and possibly bound system as previously used
as an argument in favour of the CMa galaxy
\citep{MartinezDelgado2005}.  The youth of the BP also explains
why no 1-2 Gyr red clump is evident at 8 kpc.  Fig.~\ref{fig:XYZ} also
reveals the presence of a few other BPs.  These would also be
considered a non-Galactic population, but in this case they are much
closer than the proposed distance to the CMa galaxy.

\begin{table*}
  \caption{Parameters for clusters and blue plumes. l,b are Galactic 
    longitude and latitude; Dist is the the heliocentric distance;
    X,Y,Z are Galactic Cartesian coordinates; RGC is the distance to
    the centre of the Galaxy adopting 8.5 Kpc for the Solar
    Galactocentric distance.}
\label{tab:pars}
\begin{tabular}{lccrrcccrl}
\hline
Field      & l      & b     & Dist  & Age & X      & Y     & Z     & RGC   & Constellation\\
           &($^o$)  &($^o$) & (kpc)  & (Myr) & (kpc)  &   (kpc)  &(kpc)  &(kpc)  & \\
\hline
NGC2129    & 186.55 & +0.06 & 2.19  & 10  & -0.25  & 2.18  & 0.00  & 10.68 & Gemini\\      
S203       & 210.80 & -2.56 & 8.05  & 10  & -4.12  & 6.91  & -0.36 & 15.95 & Monoceros\\   
Dolidze25  & 211.20 & -1.32 & 6.33  & 10  & -3.28  & 5.41  & -0.15 & 14.29 & Monoceros\\   
Bochum2    & 212.30 & -0.39 & 6.31  & 5   & -3.37  & 5.33  & -0.04 & 14.24 & Monoceros\\   
S285       & 213.80 & +0.61 & 7.70  & 10  & -4.28  & 6.40  & 0.08  & 15.50 & Monoceros\\   
BP2232     & 214.60 & -7.41 & 6.22  & $<$100  & -3.50  & 5.08  & -0.80 & 14.02 & Monoceros\\   
NGC2232    & 214.60 & -7.41 & 0.34  & 40  & -0.19  & 0.28  & -0.04 & 8.78  & Monoceros\\   
S289       & 218.80 & -4.55 & 9.46  & 10  & -5.91  & 7.35  & -0.75 & 16.93 & Monoceros\\   
NGC2302    & 219.28 & -3.10 & 1.37  & 40  & -0.87  & 1.06  & -0.07 & 9.60  & Monoceros\\   
NGC2302    & 219.38 & -3.10 & 7.48  & 90  & -4.74  & 5.77  & -0.40 & 15.04 & Monoceros\\   
NGC2335    & 223.62 & -1.26 & 1.79  & 79  & -1.23  & 1.30  & -0.04 & 9.87  & Canis Major\\ 
NGC2353    & 224.66 & +0.42 & 1.23  & 79  & -0.86  & 0.87  & 0.01  & 9.41  & Canis Major\\ 
BP33       & 225.40 & -3.12 & 7.69  & $<$100  & -5.47  & 5.39  & -0.42 & 14.93 & Canis Major\\ 
BP7        & 225.44 & -4.58 & 9.04  & $<$100  & -6.42  & 6.32  & -0.72 & 16.15 & Canis Major\\ 
NGC2401    & 229.67 & +1.85 & 6.31  & 20  & -4.81  & 4.08  & 0.20  & 13.47 & Puppis\\      
NGC2414    & 231.41 & +1.94 & 5.62  & 16  & -4.39  & 3.50  & 0.19  & 12.78 & Puppis\\      
BP1        & 232.33 & -7.31 & 7.69  & $<$100  & -6.04  & 4.66  & -0.98 & 14.48 & Canis Major\\ 
Bochum5    & 232.56 & 0.68  & 2.69  & 60  & -2.14  & 1.64  & 0.03  & 10.36 & Puppis\\      
S305       & 233.80 & -0.18 & 6.11  & 10  & -4.93  & 3.61  & -0.02 & 13.07 & Puppis\\      
S309       & 234.80 & -0.20 & 7.01  & 10  & -5.73  & 4.04  & -0.02 & 13.79 & Puppis\\      
BP2383     & 235.27 & -2.43 & 8.79  & $<$100  & -7.22  & 5.00  & -0.37 & 15.31 & Canis Major\\ 
NGC2384    & 235.39 & -2.42 & 2.88  & 13  & -2.37  & 1.63  & -0.12 & 10.41 & Canis Major\\ 
BP2384     & 235.39 & -2.42 & 8.79  & $<$100  & -7.23  & 4.99  & -0.37 & 15.30 & Canis Major\\ 
BP2432     & 235.48 & +1.78 & 6.00  & $<$100  & -4.94  & 3.40  & 0.19  & 12.88 & Puppis\\      
NGC2367    & 235.63 & -3.85 & 2.03  & 5   & -1.67  & 1.14  & -0.14 & 9.79  & Canis Major\\ 
BP2367     & 235.64 & -3.85 & 8.51  & $<$100  & -7.01  & 4.79  & -0.57 & 15.03 & Canis Major\\ 
NGC2362    & 238.18 & -5.55 & 1.58  & 12  & -1.34  & 0.83  & -0.15 & 9.42  & Canis Major\\ 
BP2362     & 238.18 & -5.55 & 10.81 & $<$100  & -9.14  & 5.67  & -1.05 & 16.87 & Canis Major\\ 
Trumpler7  & 238.21 & -3.34 & 2.04  & 50  & -1.73  & 1.07  & -0.12 & 9.73  & Puppis\\      
BP18       & 239.94 & -4.92 & 7.87  & $<$100  & -6.79  & 3.93  & -0.67 & 14.16 & Canis Major\\ 
Ruprecht32 & 241.50 & -0.60 & 9.68  & 10  & -8.51  & 4.62  & -0.10 & 15.64 & Puppis\\      
Haffner16  & 242.09 & +0.47 & 3.63  & 50  & -3.21  & 1.70  & 0.03  & 10.69 & Puppis\\      
Haffner19  & 243.04 & 0.52  & 5.25  & 4   & -4.68  & 2.38  & 0.05  & 11.84 & Puppis\\      
Haffner18  & 243.11 & 0.42  & 7.94  & 4   & -7.08  & 3.59  & 0.06  & 14.01 & Puppis\\      
NGC2453    & 243.35 & -0.93 & 5.25  & 40  & -4.69  & 2.35  & -0.09 & 11.83 & Puppis\\      
NGC2439    & 246.41 & -4.43 & 4.57  & 10  & -4.18  & 1.82  & -0.35 & 11.14 & Puppis\\      
BP2439     & 246.41 & -4.43 & 10.91 & $<$100  & -9.97  & 4.35  & -0.84 & 16.27 & Puppis\\      
Ruprecht35 & 246.63 & -3.24 & 5.32  & 70  & -4.88  & 2.11  & -0.30 & 11.67 & Puppis\\      
BP2533     & 247.81 & +1.29 & 6.49  & $<$100  & -6.01  & 2.45  & 0.15  & 12.49 & Puppis\\      
Ruprecht47 & 248.25 & -0.19 & 4.37  & 70  & -4.06  & 1.62  & -0.01 & 10.90 & Puppis\\      
NGC2571    & 249.10 & +3.54 & 1.38  & 50  & -1.29  & 0.49  & 0.09  & 9.08  & Puppis\\      
Ruprecht48 & 249.12 & -0.59 & 6.03  & 70  & -5.63  & 2.15  & -0.06 & 12.05 & Puppis\\      
Ruprecht55 & 250.68 & +0.76 & 4.59  & 10  & -4.33  & 1.52  & 0.06  & 10.91 & Puppis\\      
BP55       & 250.68 & +0.76 & 6.98  & $<$100  & -6.59  & 2.31  & 0.09  & 12.66 & Puppis\\      
BP2477     & 253.56 & -5.84 & 11.69 & $<$100  & -11.15 & 3.29  & -1.19 & 16.23 & Puppis\\      
NGC2547    & 264.45 & -8.53 & 0.49  & 63  & -0.48  & 0.05  & -0.07 & 8.56  & Vela\\        
Pismis8    & 265.09 & -2.59 & 2.00  & 7   & -1.99  & 0.17  & -0.09 & 8.90  & Vela\\        
NGC2910    & 275.29 & -1.17 & 1.32  & 70  & -1.31  & -0.12 & -0.03 & 8.48  & Vela\\        
\hline
\end{tabular}
\end{table*}

We now focus on the stellar groups marked with a lighter tone. These
groups are distributed between l=190$^o$ and l=270$^o$, but seem to
form an elongated structure between l = 230$^o$ and l = 250$^o$
stretching toward the outer Galaxy.  We interpret this structure as
the probable extension of the local (Orion) arm in the third quadrant.
It appears that the Orion arm stretches outward reaching and crossing
the Perseus arm.  Despite the small number of objects, a few clusters
at l$\approx$245$^o$, and one at l$\approx$185$^o$, appear to be
tracing Perseus, although less evidently at the intersection with the
local arm.

To further assess this tentative picture, we have plotted the
distribution of CO molecular clouds (kindly provided by J. May and L.
Bronfman ahead of publication and also from their previous
survey; \citealt{May1997}).  Only clouds more massive then
0.5$\times$10$^5$M$_{\odot}$ are shown.  The remarkable coincidence of
stars and clouds in the outer arm, already stressed by
us \citep{Carraro2005c}, lends further support to the interpretation of
this structure as being a spiral arm and not a tidal tail composed of
an old population. The distribution of clouds shows how between
l=180$^o$ and l=210$^o$ the arm becomes very distant from the Sun,
which is likely the reason why very distant young clusters have not
been optically identified. A good correspondence between the stellar
positions and the CO clouds is again found for the region between the
Sun and the outer arm (only data from the older survey is available
here), although not as good as the one found with the newer CO data in
the outer arm.  In particular, the Perseus arm is traced by
concentrations of CO clouds around l$\approx$220, 235 and 260$^o$. The
lower panel shows that the gas also follows the vertical trend found
for the stars in the outer arm. It is interesting to note that a
picture in some aspects not very different from the one we have just
established has been suggested more than 25 years
ago \citep{Moffat1979}.  In particular, that the local arm starts
tangent to the Carina-Sagittarius arm in the first quadrant at $l
\approx 60^o$, and then probably crosses the Perseus arm at $l\approx
240^o$, although not as clearly as in this work (likely due to the
limited depth of the older photoelectric photometry).

With these results in mind, we can address the CMa overdensity.
Although there is no consensus about the exact centre of the
overdensity, it is generally accepted that it is around $l = 240^o$,
$b = -7^o$. From Fig.~\ref{fig:XYZ} it is readily seen, both in the
gas and in the clusters, that this is the approximate direction of the
proposed extension of the Orion arm into the 3GQ. It is therefore
quite probable that we should find an overdensity of stars along this
line of sight.  Indeed, looking along the local arm right through the
middle leads us to predict $l \approx 245^o$ as the maximum density
longitude.  Interestingly, the latest estimate of the centre using red
clump stars \citep{Bellazzini2006a} is similarly at $l \approx 245^o$.
It is also worth noticing that the angle comprised by most of the
local arm extension roughly corresponds to the area claimed to contain
the overdensity. We further notice that this area should be neither
much smaller, due to the presence of the local arm, nor much larger,
due to the presence of the CO cloud complexes around $l \approx 245^o$
and $260^o$ which will limit visibility and introduce a border effect
on the observed overdensity.  In this context, we find that the CMa
overdensity is not the core of a galaxy, but simply the result of
looking along the extension of the local arm in the third quadrant.

A previous alternative explanation for the overdensity has been
proposed in which the overdensity is a signature of the Galactic
warp \citep{Momany2004}, although it does not fully explain the
Monoceros ring.  This has induced a lively debate where the warp
explanation has been partially rebated \citep{Martin2004b}. But that
counter attack, although favouring the CMa galaxy, also pointed out
that its kinematics was compatible with the velocities of a distant
arm detected in the fourth quadrant \citep{McClureGriffiths2004}. Since
this distant arm is quite likely the continuation of the Norma arm according
to model predictions \citep{Cordes2002}, the kinematic result then
further supports our picture.

Given the evidence we have presented here, our scenario is the
only one that presently accounts for, and explains, all the
observational results.  As a final remark, we note that no ad-hoc new
spiral arms had to be introduced.  All the spiral features we evoke
are simply previously unclear extensions of well known arms whose
existence has been repeatedly established during the last few decades.
In this context, the presence of spiral arms in a region of a spiral
galaxy where they are are supposed to be, but have not been detected
before, is a more natural explanation of the CMa phenomena than a
cannibalised galaxy in a nearly co-planar orbit, how exciting it might
be.

\section*{Acknowledgments}
We thank J. May and L. Bronfman for helpful discussions and for
providing results ahead of publication and P.G.  Ferreira for language
editing. A.M. acknowledges grant SFRH/BPD/19105/2004 from {\it FCT}
(Portugal).  G.C. acknowledges {\it Fundaci\'on Andes}. R.A.V., G.C.,
G.B. and E.G.  acknowledge {\it Programa Cient\'ifico-Tecnol\'ogico
  Argentino-Italiano SECYT-MAE IT/PA03-UIII/077-2004-2005}. This study
made use of Simbad and WEBDA databases.


\label{lastpage}
\end{document}